# Chapter 1

## KEYWORDS, ETC.

```
(The first two pages only resemble the abtract infos and are
needed to ensure proper layout of the rest of the article.
Bernhard)
--------
Name(s) and affiliation (s):
Andy Evans
University of York, York, UK
andye@cs.york.ac.uk

Robert France
Colorado State University, Colorado, US
france@cs.colostate.edu

Kevin Lano
Imperial College, London, UK
kcl@doc.ic.ac.uk

Bernhard Rumpe
Software \& Systems Engineering
Munich University of Technology, Munich, Germany
rumpe@in.tum.de

Title:
Meta-Modelling Semantics of UML

Index items:
UML, Meta-model, Theory, Formalization, OCL, Semantics,
Generalisation
```

```
Abstract:
---------
The Unified Modelling Language is emerging as a de-facto standard for
modelling object-oriented systems. However, the semantics document that
a part of the standard definition primarily provides a description
of the language's syntax and well-formedness rules. The meaning of the
language, which is mainly described in English, is too informal
```

and unstructured to provide a foundation for developing formal analysis and development techniques. This paper outlines a formalisation strategy for making precise the core semantics of UML. This is acheived by strengthening the denotational semantics of the existing UML semantics. To illustrate the approach, the semantics of generalization/specialization
are made precise.

# Chapter 4

## META-MODELLING SEMANTICS OF UML

**Andy Evans**
*University of York, UK*
*andye@cs.york.ac.uk*

**Kevin Lano**
*Imperial College, UK*
*kcl@doc.ic.ac.uk*

**Robert France**
*Colorado State University, US*
*france@cs.colostate.edu*

**Bernhard Rumpe**
*Munich University of Technology*
*Germany*
*rumpe@in.tum.de*

**Abstract**     The Unified Modelling Language is emerging as a de-facto standard for modelling object-oriented systems. However, the semantics document that a part of the standard definition primarily provides a description of the language's syntax and well-formedness rules. The meaning of the language, which is mainly described in English, is too informal and unstructured to provide a foundation for developing formal analysis and development techniques. This paper outlines a formalisation strategy for making precise the core semantics of UML. This is achieved by strengthening the denotational semantics of the existing UML metamodel. To illustrate the approach, the semantics of generalization/specialization are made precise.

## 1.     INTRODUCTION

The Unified Modeling Language (UML) [BRJ98, RJB99] is rapidly becoming a de-facto language for modelling object-oriented systems. An important aspect of the language is the recognition by its authors of the need to provide a precise description of its semantics. Their intention is that this should act as an unambiguous description of the language, whilst also permitting extensibility so that it may adapt to future changes in object-oriented analysis and design. This has resulted in a Semantics Document [OMG99], which is presently being managed by the Object Management Group, and forms an important part of the language's standard definition.

The UML semantics is described using a meta-model that is presented in terms of three views: the abstract syntax, well-formedness rules, and modelling element semantics. The abstract syntax is expressed using a subset of UML static modelling notations. The abstract syntax model is supported by natural language descriptions of the syntactic structure of UML constructs. The well-formedness rules are expressed in the *Object Constraint Language* (OCL) and the semantics of modelling elements are described in natural language. The advantage of using the meta-modelling approach is that it is accessible to anybody who understands UML. Furthermore, the use of object-oriented modelling techniques helps make the model more intuitively understandable.

A potential advantage of providing a precise semantics for UML is that many of the benefits of using a formal language such as Z [S92] or Spectrum [BFG$^+$93] might be transferable to UML. Some of the major benefits of having a precise semantics for UML are given below:

> **Clarity**: The formally stated semantics can act as a point of reference to resolve disagreements over intended interpretation and to clear up confusion over the precise meaning of a construct.
>
> **Equivalence and Consistency**: A precise semantics provides an unambiguous basis from which to compare and contrast the UML with other techniques and notations, and for ensuring consistency between its different components.
>
> **Extendibility**: The soundness of extensions to the UML can be verified (as encouraged by the UML authors).
>
> **Refinement**: The correctness of design steps in the UML can be verified and precisely documented. In particular, a properly developed semantics supports the development of design transformations, in which a more abstract model is diagrammatically transformed into an implementation model.
>
> **Proof**: Justified proofs and rigorous analysis of important properties of a system described in UML require a precise semantics in order to determine their correctness.

Unfortunately, the current UML semantics are not sufficiently formal to realise these benefits. Although much of the syntax of the language has been defined, and some static semantics given, its semantics are mostly described using lengthy paragraphs of often ambiguous informal English, or are missing entirely. Furthermore, limited consideration has been paid to important issues such as proof, compositionality and rigorous development. A further problem is the extensive scope of the language, all of which must be dealt with before the language is completely defined.

This chapter describes work being carried out by the precise UML (pUML) group and documented in [PUML99, FELR98, EFLR98]. PUML is an international group of researchers and practitioners who share the goal of developing UML as a precise (formal) modelling language, thereby enabling it to be used in a formal manner. This chapter reports on work being carried out by the group to strengthen the existing semantics of UML. In Section 2., a formalisation strategy is described (developed through the experiences of the group) that aims to make precise the existing UML semantics. A core UML semantics model is identified in Section 3. as a first step towards achieving this goal. Section 4. then describes how the formalisation strategy has been applied to the development of a precise understanding of a small yet interesting part of the UML semantics - generalization/specialization hierarchies. Finally, the paper concludes with a brief overview of some future directions of the group's work.

## 2.     FORMALISATION STRATEGY

In order to implement the pUML approach it is necessary to develop a strategy for formalising the UML. This is intended to act as a step by step guide to the formalisation process, thus permitting a more rigorous and traceable work program.

In developing a formalisation strategy for UML it has been necessary to consider the following questions:

1. Is the meta-modelling approach used in the current UML semantics suitable for assigning a precise semantics to UML?

2. Should the existing UML semantics be used as a foundation for developing a precise semantics for UML?

3. Given the large scope of UML, which parts should be formalised first?

**Suitability of meta-modelling**

There are many approaches used to assign semantics to languages. One of the best known (and most popular) is the denotational approach (for an in-depth discussion see [S86]). The denotational approach assigns semantics to a language by giving a mapping from its syntactical representation to a meaning, called a denotation. A denotation is usually a well-defined mathematical value, such as a number or a set. Typically, functions are used to define mappings between syntax and denotations. For example, the meaning of a simple language for adding and subtracting natural numbers might be described in terms of two functions, add and subtract, and the result of each would be a single integer value.

The use of a language to give a 'meta-circular' description of its own denotational semantics is well known in Computer Science. For example, the specification language Z has been given a meta-circular semantics using a simple subset of Z [S92]. Unfortunately, the meta-modelling approach opens itself to the criticism that it doesn't really define anything. Informally, if a reader does not understand UML, then it is unlikely that they will understand the meaning of UML when written in UML.

The justification given for using meta-modelling in these contexts is that, in principle at least, it should be possible to give a formal interpretation to a meta-description in terms of a more basic language such as predicate logic. This argument can also be applied to UML, as it seems likely that it can be given a more fundamental interpretation in terms of sets and predicate logic. Indeed, a significant amount of work has already been done to describe the semantics of UML class diagrams and OCL like expressions [BR98] in Z. There is also an important pragmatic reason for choosing UML to describe the denotational semantics of UML: Because UML is designed to provide an intuitive means for constructing models, using UML to help better understand UML is likely to be a useful way of testing the expressiveness and power of UML as a modelling language.

Given that UML can be used to describe its own semantics, how should these semantics be presented in order to emphasise the denotational approach? As described in the introduction, the current UML semantics already makes a distinction between syntax and semantics (as in the denotational approach). However, it mainly uses English prose to describe the semantic part. The pUML approach advances this work by using associations (and constraints on associations) to map syntactical elements to their denotations. This approach has also been used in the UML semantics to a limited extent. For example, associations are described by the set of possible object links they are associated with. The distinguishing feature of the pUML approach is its emphasis on obtaining *precise* denotational descriptions of a much wider selection of UML modelling elements.

### Working with the standard

Assuming that a meta-modelling approach is adopted to describe the UML semantics, two approaches to developing a precise semantics can be adopted. The first approach is to ignore the existing semantics documentation and develop a new model. This has the advantage that the modeller is completely free to develop a semantics that is appropriate to their needs. For example, greater emphasis might be placed on obtaining a simple semantic model, or one that will readily support a particular proof technique.

The second approach is to adopt the existing semantics as a foundation from which a precise semantics can be obtained. Some good reasons for adopting this approach are as follows:

1. It recognises that considerable time and effort has been invested in the development of the existing UML semantics. It cannot be expected that a radically different semantic proposal will be incorporated in new versions.

2. Without working within the constraints of the existing semantics it is easy to develop models that are incompatible with the standard or omit important aspects of it.

An important aspect of the pUML approach is its aim of eventually contributing to the emerging standard. Therefore, it is the second approach that has been adopted. This is why the remainder of the paper will focus on developing an approach to incrementally clarifying the *existing* semantics of UML.

### Clarifying a core semantics

To cope with the large scope of the UML it is natural to concentrate on essential concepts of the language to build a clear and precise foundation as a basis for formalisation. Therefore, the approach taken in the group's work is to concentrate on identifying and formalising a core semantic model for UML before tackling other features of the language. This has a number of advantages: firstly, it makes the formalisation task more manageable; secondly, a more precise core will act as a foundation for understanding the semantics of the remainder of the language. This is useful in the case of the many diagrammatical notations supported by UML, as each diagram's semantics can be defined as a particular 'view' of the core model semantics. For example, the meaning of an interaction diagram should be understandable in terms of a subset of the behavioural semantics of the core.

**Formalisation strategy**

The formalisation strategy consists of the following steps:

1. Identify the core elements of the existing UML semantics.

2. Iteratively examine the core elements, seeking to verify their completeness. Here, completeness is achieved when: (1) the modelling element has a precise syntax, (2) is well-formed, and (3) has a precise denotation in terms of some fundamental aspect of the core semantic model.

3. Use formal techniques to gain better insight into the existing definitions as shown in [FELR98, EFLR98].

4. Where in-completeness is identified, we attempt to address it in a number of ways, depending on the type of omission found.

    Model strengthening - this is necessary where the meaning of a model element is not fully described in the meta-model. The omission is fixed by strengthening the relationship between the model element and its denotation.

    Model extension - in certain cases it is necessary to extend the meta-model to incorporate new denotational relationships. This occurs when no meaning has been assigned to a particular model element, and it cannot be derived by constraints on existing associations. For example, this is necessary in the case of *Operation* and *Method*, where the meaning of a method is defined in terms of a *procedureExpression* and Operation is given no abstract meaning at all.

    Model simplification - in some cases, aspects of the model are surplus to needs, in which case we aim to show how they can be omitted or simplified without compromising the existing semantics.

5. Feed the results back into the UML meta-model, with the aim of clarifying the semantics of a core part of the UML.

6. Disseminate to interested parties for feedback.

Finally, it is important to consider how the notion of *proof* can be represented in the semantic model. This is essential if techniques are to be developed for analysing properties of UML models. Such analysis is required to establish the presence of desired properties in models [E98]. The need to establish properties can arise out of the need to establish that models adhere to requirements or out of challenges posed by reviewers of the models. Proof is also important in understanding properties of model transformations in which a system is progressively refined to an implementation [BHH$^+$97].

## 3. THE CORE SEMANTICS MODEL

The question of what should form a core precise semantics for UML is already partially answered in the UML semantics document. It identifies a 'Core Package - Relationships' package and a number of 'Common Behaviour' packages. The Core Relationship package defines a set of modelling elements that are common to all UML diagrams, such as ModelElement, Relationship, Classifier, Association and Generalization. However, it only describes their syntax. The Common Behavior (Instances and Links) package gives a partial denotational meaning to the model elements in the core package. For instance, it describes an association between Classifier and Instance. This establishes the connection between the representation of a Classifier and its meaning, which is a collection of instances. The meaning of Association (a collection of Object Links) is also given, along with a connection between Association roles and Attribute values.

To illustrate the scope, and to show the potential for realising a compact core semantics, the relevant class diagrams of the two models are shown in the Figures 4.1 and 4.2. Well-formedness rules are omitted for brevity.

An appropriate starting point for a formalisation is to consider these two models in isolation, with the aim of improving the rigor with which the syntax of UML model elements are associated with (or mapped to) their denotations.

## 4. FILLING THE SEMANTIC GAP

In this section, we illustrate how the pUML formalisation approach has been applied to a small part of the core model. The modelling concept that will be investigated is generalization/specialization.

### 4.1 DESCRIPTION

In UML, a generalization is defined as "a taxonomic relationship between a more general element and a more specific element", where "the more specific element is fully consistent with the more general element" [OMG99], page 2-34 (it has all of its properties, members, and relationships) and may contain additional information.

Closely related to the UML meaning of generalization is the notion of direct and indirect instances: This is alluded to in the meta-model as the requirement that "an instance is an indirect instance of ... any of its ancestors" [OMG99], page 2-56.

UML also places standard constraints on subclasses. The default constraint is that a set of generalizations are disjoint, i.e. " (an) instance may have no more than one of the given children as a type of the instance" [OMG99], page 2-35. Abstract classes enforce a further constraint, which implies that no instance can be a direct instance of an abstract class.

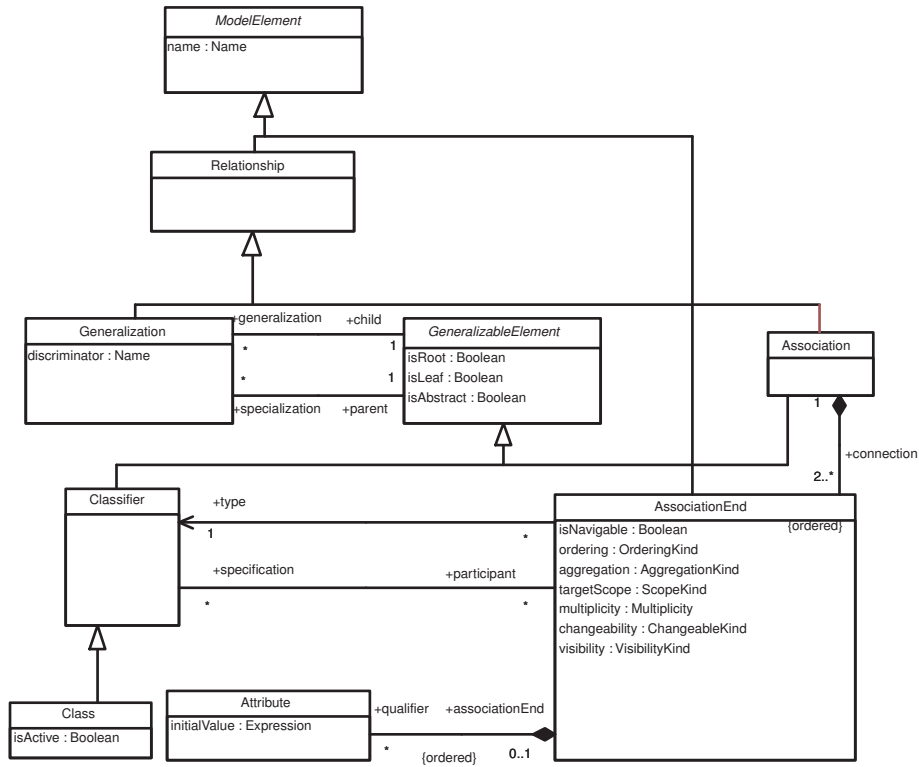

*Figure 4.1*   Fragment of the core relationships package

We now examine whether these properties are adequately specified in the UML semantics document. In this paper, we will only consider properties that relate to Classifiers: the UML term for any model element that describes behavioural and structural features. Classes are a typical specialisation of Classifiers.

## 4.2    EXISTING FORMAL DEFINITIONS

France et al. [BR98] have defined a formal model of generalization that fits very well with that adopted in UML. Classes are denoted by a set of object references, where each reference maps to a set of attribute values and operations. generalization implies inheritance of attributes and operations from parent classes (as expected). In addition, class denotations are used to formalise the meaning of direct and indirect instances, disjoint and abstract classes. This is achieved by constraining the sets of objects assigned to classes in different ways depending on the roles the classes play in a particular generalization hierarchy. For example, assume that $ai$ is the set of object references belonging to the class $a$, and $b$ and $c$ are subclasses of $a$. Because instances of $b$ and $c$ are also indirect instances of $a$, it is required that $bi \subseteq ai$ and $ci \subseteq ai$, where $bi$ and $ci$ are the set of object references of $b$ and $c$. Thus, a direct instance of $b$ or $c$ must also be an *indirect* instance of $a$. A direct instance is also distinguishable from an indirect instance if there does not exist a specialised class of which it is also an instance.

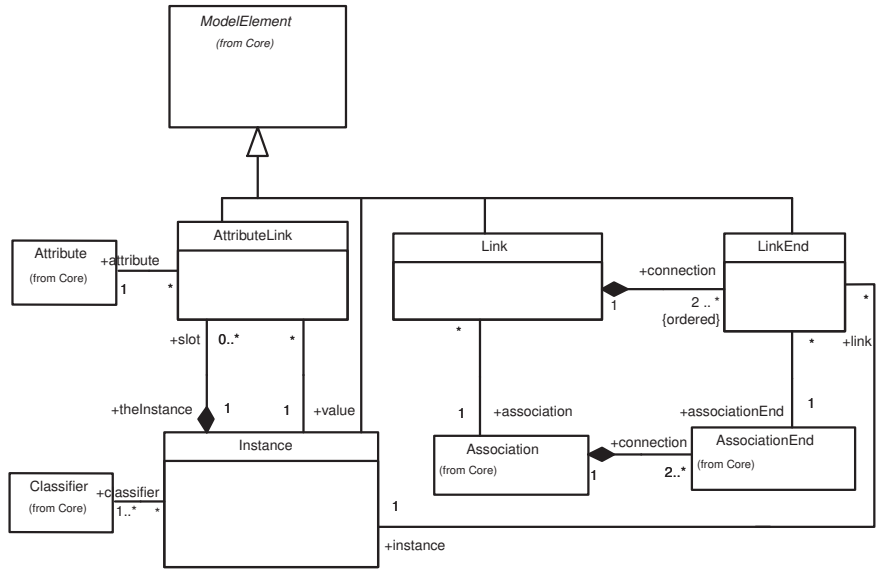

*Figure 4.2* Fragment of the common behaviour package

This model also enables constraints on generalizations to be elegantly formalised in terms of simple constraints on sets of object references. In the case of the standard 'disjoint' constraint on subclasses, the following must hold: $bi \cap ci = \emptyset$, i.e. there can be no instances belonging to both subclasses. For an abstract class, this constraint is further strengthened by requiring that $bi$ and $ci$ partition $ai$. In other words, there can be no instances of $a$, which are not instances of $b$ or $c$. Formally, this is expressed by the constraint: $bi \cup ci = ai$.

We will adopt this model in order to assign a precise denotational meaning to generalization/specialization.

## 4.3   SYNTAX AND WELL-FORMEDNESS

The abstract syntax of generalization/specialization is described by the meta-model fragment in Figure 4.3 of the core relationships package:

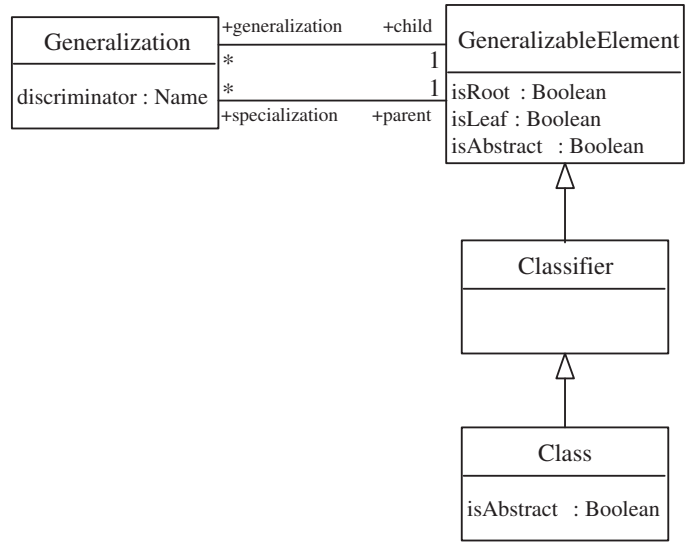

*Figure 4.3*   Meta-model fragment of Generalization/Specialization

The most important well-formedness rule which applies to this model element, and is not already ensured by the class diagram, is that circular inheritance is not allowed. Assuming `allParents` defines the transitive closure of the relationship induced by `self.generalization.parent`, which happens to be the set of all ancestors, then it must hold that:

```
context GeneralizableElement
not self.allParents -> includes(self)
```

## 4.4   SEMANTICS

The completeness of the semantic formalisation vs. the desired properties of generalization is now examined. We concentrate on determining whether the following properties of generalization are captured in the meta-model:

- instance identity and conformance.

- direct and indirect instantiation of classifiers.

- disjoint and overlapping constraints on sub-classifiers.

- abstract classes.

As noted in Section 3., the UML meta-model already describes a denotational relationship between Classifier and Instance. The meta-model fragment in Figure 4.4 describes this relationship.

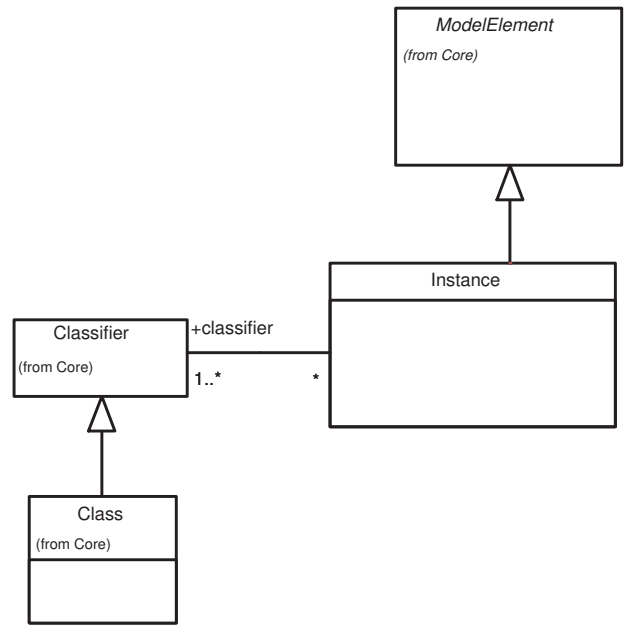

*Figure 4.4*   Meta-model fragment for Class and Instance relationship

However, unlike the formal model described above, the UML meta-model does not describe the constraints that generalization implies on this relationship. For example, an *Instance* can be an instance of many classifiers, yet there are no constraints that the classifiers are related. Thus, the meta-model must be strengthened with additional constraints on the relationship between model elements and their denotations.

## 4.5    MODEL STRENGTHENING

The first aspect of the model to be strengthened relates to the meaning of indirect instances. As stated in Section 4.1, an instance of a classifier is also an *indirect* instance of its parent classifiers. This property, which we term as 'instance conformance' can be precisely stated by placing an additional constraint on the relationship between the instances of a classifier and the instances belong to the classifier's parents. It is specified as follows:

```
context c :   Classifier
invariant
   c.generalization.parent -> forall(s :  Classifier |
       s.instance -> includesAll(c.instance))
```

This states that the instances of any Classifier, c, are a subset of those belonging to the instances of its parents.

*4.5.1  Direct instances.*  Given the above property, it is now possible to precisely describe the meaning of a direct instance:

```
context i :  Instance
isDirectInstanceOf(c :  Classifier) :  Boolean
isDirectInstanceOf(c) =
                c.allParents -> union(Set(c)) = i.classifier
```

A direct instance directly instantiates a single class and indirectly instantiates all its parents. This definition is in fact a more precise description of the OCL operation `oclIsTypeOf`, i.e.

```
context i :  Instance
oclIsTypeOf(c :  Classifier) :  Boolean
oclIsTypeOf(c) = i.isDirectInstanceOf(c)
```

A similar operation can be used to assign a precise meaning to the OCL operation `oclIsKindOf`:

```
context i :  Instance
oclIsKindOf(c :  Classifier) :  Boolean
oclIsKindOf(c) = i.oclIsTypeOf(c) or
                c.allSupertypes ->
                    exists(s :  Classifier | i.oclIsTypeOf(s))
```

Finally, an OCL operation which returns the Classifier from which an instance is directly instantiated from can be defined:

```
context i :  Instance
direct :  Classifier
direct = i.classifier -> select(c | i.isDirectInstanceOf(c))
```

*4.5.2  Indirect instances.*  Once the meaning of a direct instance is defined, it is straightforward to obtain an OCL operation that returns all the Classifiers that an instance indirectly instantiates.

```
context i :  Instance
indirect :  Set(Classifier) :
indirect = i.classifier - Set(i.direct)
```

The set of indirect classes is the difference of the set of all classifiers instantiated by the instance and the direct classifier.

*4.5.3  Instance identity.*  Unfortunately, the above constraints do not guarantee that every instance is a direct or indirect instance of a related classifier. For example, consider two classifiers that are not related by generalization/specialization. The current UML semantics do not rule-out the possibility of an instance being instantiated by both classifiers.

Thus, an additional constraint must be added in order to rule out the possibility of an instance being instantiated from two or more un-related classes. This is the unique identity constraint:

```
context i :   Instance
invariant
    i.classifier = i.direct -> union(i.indirect)
```

This states that the *only* classifiers that an object can be instantiated from are either the classifier that it is directly instantiated from or those that it is indirectly instantiated from.

*4.5.4   Disjoint subclasses.*   Once direct and indirect instances are formalised, it is possible to give a precise description to the meaning of constraints on generalizations (for example the disjoint constraint).

The disjoint constraint can be formalised as follows:

```
context c :   Classifier
invariant
   c.specialization.child -> forall(i,j :  Classifier |
       i <> j implies i.instance ->
           intersection(j.instance) -> isEmpty)
```

This states that for any pair of direct subclasses of a class, `i` and `j`, the set of instances of `i` will be disjoint from the set of instances of `j`.

*4.5.5   Abstract classes.*   Finally, the following OCL constraint formalises the required property of an abstract class that it can not be directly instantiated:

```
context c :   Classifier
invariant
   c.isAbstract implies
        c.specialization.child.instance -> asSet = c.instance
```

Note, the result of the `specialization.child` path is a bag of instances belonging to each subclass of `c`. Applying the asSet operation results in a set of instances. Equating this to to the instances of `c` implies that all the instances of `c` are covered by the instances of its subclasses. This, in conjunction with the disjoint property above, implies the required partition of instances.

## 4.6   MODEL EXTENSION

The above definition of the 'disjoint' constraint is adequate provided that it applies across all generalizations, and indeed this is the default assumption in UML. However, UML also permits overlapping constraints to be applied across subclasses as shown in Figure 4.5.

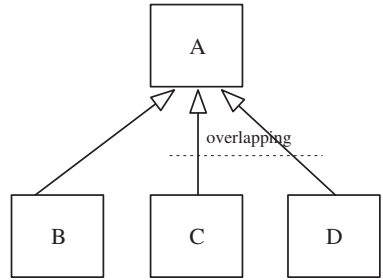

*Figure 4.5* Partly overlapping subclasses

Here, instances of C and D may overlap, but they must be disjoint from instances of B (the default disjoint constraint still exists between B and C and B and D). Thus, the overlapping constraint is viewed as overriding the existing default constraint.

Unfortunately, overlapping constraints are not explicitly encoded in the existing semantics. Therefore, it is necessary to extend the meta-model with an explicit overlapping constraint in order to be able to formalise its meaning. This is shown in Figure 4.6.

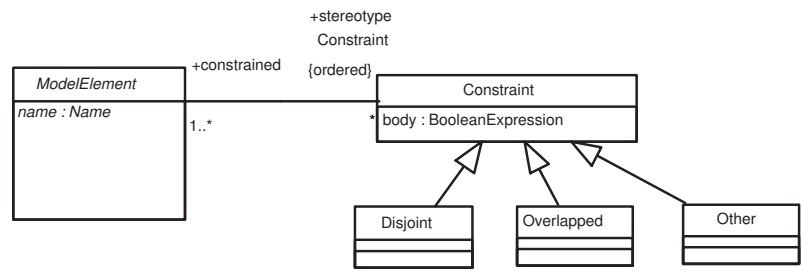

*Figure 4.6* Fragment of the meta-model with extended Constraint

Here, overlapping constraints are modelled as a subclass of *Constraint*. Because overlapping constraints must be applied across more than one subclass, the following additional well-formedness rule must be added:

```
context o :  Overlapping
invariant
    o.constrained -> size > 1
```

An improved version of the disjoint constraint can now be given:

```
context c :  Classifier
invariant
```

```
    c.specialization -> forall(i,j : Generalization |
       (i <> j and
       not (i.hasSameOverlappingConstraint(j)))
       implies i.child.instance ->
           intersection(j.child.instance) -> isEmpty)
```

This states that the instances of two or more generalizations are disjoint unless they overlap. Note that the same overlapping constraint must be applied to the generalizations.

The operation `hasSameOverlappingConstraint` is defined as follows:

```
context i :   Generalization
hasSameOverlappingConstraint(j :  Generalization) :   Boolean

hasSameOverlappingConstraint(j) =
    ((i.stereotypeConstraint -> asSet) ->
        intersection(j.stereotypeConstraint -> asSet) ->
            exists(c :  Constraint | c.oclType = Overlapping))
```

This operation is true if a pair of generalizations share the same overlapping constraint.

This completes the formalisation examples. Although not complete, they indicate the benefits of adopting a denotational emphasis in modelling the UML semantics. In particular, they have provided a much improved understanding of some important aspects of UML. They have also provided a foundation from which to clarify many other aspects of the language, for example, the meaning of the OCL operations oclIsKindOf and oclIsTypeOf.

## 5. CONCLUSION

This paper has described ongoing work by members of the precise UML group, who are seeking to develop UML as a precise modelling language. By applying previous knowledge and experience in formalising OO concepts and semantic models, it has been shown how important aspects of the current UML semantics can be clarified and made more precise. A formalisation strategy was also described, with the aim that it will act as a template for exploring further features of UML and for developing new proof systems for the standard language.

In the longer term, our intention is to give a semantics to the complete notation set, by mapping into the core, extending the core only when there is not already a concept which suffices. Of course one role of semantics is to clarify and remove ambiguities from the notation. Therefore we will not be surprised if we find that the notation needs to be adjusted or the informal semantics rewritten. However, we will be able to provide a tightly argued, semantically-based recommendation for any change deemed necessary.

Some consideration also needs to be given to quality insurance. There are at least three approaches we have identified:

1. peer review and inspection

   2. acceptance tests

   3. tool-based testing environment

So far the only feedback has come from 1. Since a meta-model is itself a model, acceptance tests could be devised as they would be for any model. Perhaps "testing" a model is a novel concept: it at least comprises devising object diagrams, snapshots, that the model must/must-not accept. Better than a list of acceptance tests on paper would be a tool embodying the meta-model, that allowed arbitrary snapshots to be checked against it.

Finally, we re-iterate the factors driving the work outlined in this paper. Given the UML's intended role as a modelling notation standard, it is imperative that it has a well-founded semantics. Only once such a semantics is provided can the UML be used as a rigorous modelling technique. Moreover, the formalisation of UML constructs is an important step towards gaining a deeper understanding of OO concepts in general, which in turn can lead to the more mature use of OO technologies. These insights will be gained by exploring consequences of particular interpretations, and by observing the effects of relaxing and/or tightening constraints on the UML semantic model.

## Acknowledgments

This material is partially based upon work supported by: the National Science Foundation under Grant No. CCR-9803491; the Bayerische Forschungsstiftung under the FORSOFT research consortium and the DFG under the Leibnizpreis program, and the Laboraturio de Methodos Formais of the Departamento de Informatica of Pontificia Universidade Catolica do Rio de Janeiro.

## References

[BFG$^+$93] M. Broy, C. Facchi, R. Grosu, R. Hettler, H. Hußmann, D. Nazareth, F. Regensburger, O. Slotosch, and K. Stølen. The Requirement and Design Specification Language SPECTRUM, An Informal Introduction, Version 1.0, Part 1. Technical Report TUM-I9312, Technische Universität München, 1993.

[BHH$^+$97] Ruth Breu, Ursula Hinkel, Christoph Hofmann, Cornel Klein, Barbara Paech, Bernhard Rumpe, and Veronika Thurner. Towards a formalization of the unified modeling language. In Satoshi Matsuoka Mehmet Aksit, editor, *ECOOP'97 Proceedings*. Springer Verlag, LNCS 1241, 1997.

[BR98] J-M. Bruel and R.B.France. Transforming UML models to formal specifications. In *UML'98 - Beyond the notation*, LNCS 1618. Springer-Verlag, 1998.

[BRJ98] G. Booch, J. Rumbaugh, and I. Jacobson. *The Unified Modeling Language User Guide*. Addison-Wesley, 1998.

[EFLR98] Andy Evans, Robert France, Kevin Lano, and Bernhard Rumpe. Developing the UML as a formal modelling notation. In Jean Bezivin and Pierre-Allain Muller, editors, *UML'98 Proceedings*. Springer-Verlag, LNCS 1618, 1998.

[E98]     A. S. Evans. Reasoning with UML class diagrams. In *WIFT'98*. IEEE Press, 1998.
[FELR98]  R. France, A. Evans, K. Lano, and B. Rumpe. The UML as a formal modeling notation. *Computer Standards & Interfaces*, 19, 1998.
[OMG99]   Object Management Group. OMG Unified Modeling Language Specification, version 1.3r2. found at: http://www.rational.org/uml. 1999.
[PUML99]  The pUML Group. The precise UML web site: http://www.cs.york.ac.uk/puml. 1999.
[RJB99]   J. Rumbaugh, I. Jacobson, and G. Booch. *The Unified Modeling Language Reference Manual*. Addison-Wesley, 1999.
[S86]     D. A. Schmidt. *Denotational Semantics: A Methodology for Language Development*. Allyn and Bacon, 1986.
[S92]     J.M. Spivey. *The Z Reference Manual, 2nd Edition*. Prentice Hall, 1992.

## About the Authors

**Andy Evans** has taught and researched in the area of formal methods and their application to object-oriented and real-time systems. He is co-founder of the precise UML group and forthcoming co-chair of UML'2000. He has authored and co-authored papers on formalising UML and object-oriented standards.

**Robert France** is currently an Associate Professor in the Computer Science Department at Colorado State University. Currently, his primary research activities revolve around the formalization of object-oriented modeling concepts and the development of rigorous software development techniques.

**Kevin Lano** has carried out research and development using formal methods both in industry and academia. He is the author of "Formal Object-oriented Development" (Springer, 1995) and "The B Language and Method" (Springer, 1996). His current research is on the integration of formal methods and safety analysis techniques, and on the formalisation of UML.

**Bernhard Rumpe** has taught and supervised research in the area of object-oriented modelling and programming, formal methods and embedded systems. His work includes refinement and composition techniques for structural as well as behavioral notations, methodical guidelines, and the development of formalisation approaches for UML. He co-authored and co-edited three books. He is program chair of UML'99.